\documentclass[prb,reprint]{revtex4-1}
\usepackage{graphicx}
\usepackage{bm,braket,color,ulem}
\usepackage{amssymb}

\begin{document}

\title{New method for torque magnetometry using a commercially available membrane-type surface-stress sensor}

\author{Hideyuki Takahashi$^{1}$\thanks{E-mail address: hide.takahashi@crystal.kobe-u.ac.jp}, Kento Ishimura$^2$, Tsubasa Okamoto$^2$, Eiji Ohmichi$^{2}$, Hitoshi Ohta$^{3}$}
\affiliation{$^{1}$Organization for Advanced and Integrated Research, Kobe University, 1-1, Rokkodai, Nada, Kobe 657-8501, Japan\\
$^{2}$Graduate School of Science, Kobe University, 1-1 Rokkodai-cho, Nada, Kobe 657-8501, Japan\\
$^{3}$Molecular Photoscience Research Center, Kobe University, 1-1 Rokkodai-cho, Nada, Kobe 657-8501, Japan}

\date{\today}

\begin{abstract}
We present a new method for torque magnetometry by using a commercially available membrane-type surface-stress sensor (MSS). This sensor has a silicon membrane supported by four beams in which piezoresistive paths are integrated. Although originally developed as a gas sensor, it can be used for torque measurement by modifying its on-chip aluminum interconnections. We demonstrate the magnetic-torque measurement of submillimeter-sized crystals at a low temperature and in strong magnetic fields. This MSS can observe de-Haas-van-Alphen oscillation, which confirms that it can be an alternative tool for self-sensitive microcantilevers.

\end{abstract}

\maketitle
Self-sensitive microcantilevers have proven to be useful tools in torque magnetometry~\cite{Rossel1996, Naughton1997}.
They change the electrical properties of an integrated circuit, such as piezoresistance and capacitance, depending on the cantilever displacement. Because of their high force-sensitivity, the sample mass can be greatly reduced from conventional magnetization measurement techniques. Nowadays, their application is not limited to the study of magnetic materials and extends to many fields of condensed matter research; for example, using microcantilevers for de-Haas-van-Alphen oscillation measurement is advantageous as they can be used at a low temperature and even in strong pulsed magnetic fields~\cite{Ohmichi2002}; their combination with rotated magnetic fields revealed a small susceptibility anisotropy induced by electron nematicity in strongly correlated materials~\cite{Okazaki2011,Kasahara2012}.
In addition, their application to magnetic resonance spectroscopy is also an interesting topic. In our previous studies, we have demonstrated torque-detected electron spin resonance (ESR) spectroscopy in the terahertz region~\cite{Ohmichi2008, Ohmichi2009, HT2015}. 

However, the number of commercially available self-sensitive microcantilevers is so limited that their supply has been unstable. In particular, the situation became more serious after the production of PRC series microcantilevers (Hitachi High-Technologies Corp.), which once were the most popular commercially available piezoresistive microcantilevers, was stopped. Currently, the development of an alternative technique is an urgent issue. In this study, we present a new method for torque magnetometry by using a commercially available membrane-type surface-stress sensor (MSS), which was recently released by NANOSENSORS$^\mathrm{TM}$~\cite{Genki2011, Genki2012, Nanosensors}. 
Although it was originally developed as a gas sensor, the MSS can easily be diverted for use in torque magnetometry.

\begin{figure}[tb]
	\begin{center}
		\includegraphics[width=0.8\hsize]{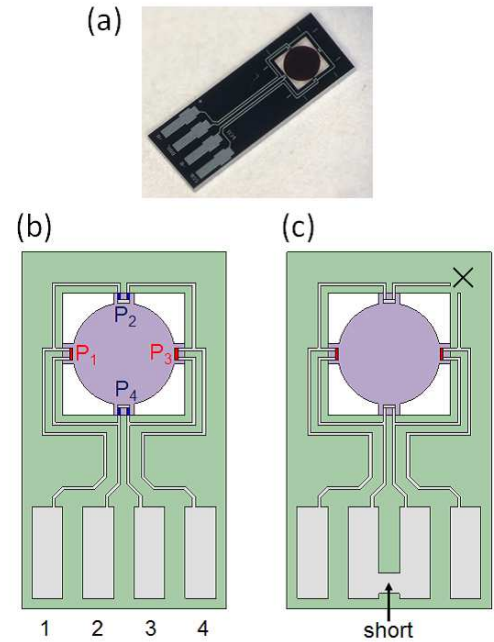}
	\end{center}
\caption{(a) Photograph of commercially available MSS chip. (b) Schematic of the wiring pattern and the pad numbering. The red and blue paths show the pairs of piezoresistors having the same shapes and resitance values. (c) Modification of wiring pattern for magnetic torque measurement. The symbol ($\times$) denotes the area where the aluminum line was scraped.}
\label{fig1}
\end{figure}

Figure~\ref{fig1}(a) shows the commercially available MSS chip (SD-MSS-1K), whose dimensions are $5.5\times 2.0\times 0.3\ \mathrm{mm}^3$~\cite{Nanosensors}.
The circular silicon membrane has a diameter and a thickness of $1\ \mathrm{mm}$ and $5.2\ \mathrm{\mu m}$, respectively.  It is supported by four beams that connect it with the outer frame. 
To detect the deformation of membrane, the piezoresistive paths P$_i$ ($i=$1-4) with resistance values of $R_i$ are fabricated on each beam by boron diffusion, and they form a Wheatstone bridge network on the chip. 
Typical values of $R_i$ are 5-12 $\mathrm{k\Omega}$ at a room temperature, which decreases at 4.2 $\mathrm{K}$ by $\sim20\ \%$.
The pairs of resistors facing across the membrane, (P$_1$, P$_3$) and (P$_2$, P$_4$), have the same shapes (see fig~\ref{fig1}(b)).
Consequently, $R_1\simeq R_3$ and $R_2\simeq R_4$, whereas $R_1\neq R_2$. 
Four electric pads are printed to connect the on-chip bridge circuit and external electronic devices.
The chip is compatible with a commercially available flexible printed circuit (FPC)/flexible flat cable (FFC) connector with a 0.5-mm pitch (54550-0471, Molex, Inc.).
We soldered an FPC/FFC connector on a pitch-conversion adapter and then placed it on an 8-pin dual in-line package integrated circuit socket. These accessories make handling easier and reduce the preparation time.

When the MSS is used as a gas sensor, the membrane is coated with a receptor layer to adsorb target molecules~\cite{Imamura2016}. 
The adsorbate creates stress in the membrane surface, which results in isotropic circular deformation. The sensor output can be obtained by measuring the voltage difference between Pads 1 and 3, which can be written as follows:
\begin{equation}
\Delta V=\frac{R_2 \Delta R_4+R_4 \Delta R_2-R_1 \Delta R_3-R_3 \Delta R_1}{R_1 + R_2 +R_3 +R_4}I,
\end{equation}
where $I$ is the bias current and $\Delta R_i$ are the change in $R_i$. 
The geometrical shapes of (P$_1$, P$_3$) and (P$_2$, P$_4$) were designed such that the piezoresistance shows a positive (negative) and a negative (positive) change, respectively, against compressive (tensile) stress.
Therefore, the changes at all resistors effectively contribute to enhance the bridge output.

\begin{figure}[tb]
	\begin{center}
		\includegraphics[width=0.9\hsize]{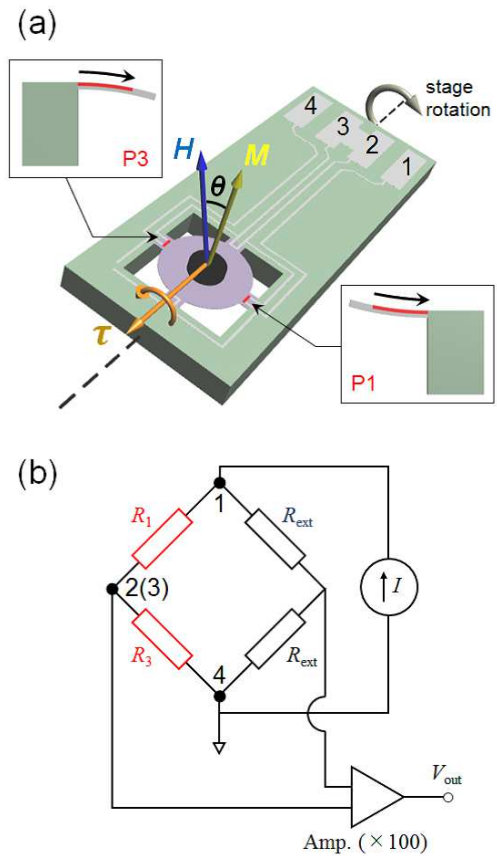}
	\end{center}
\caption{(a) Relationship between the rotation axis and the active piezoresistors in the magnetic-torque measurement. The blue and yellow arrows indicate the magnetic field direction and the normal direction to the membrane surface, respectively. The insets show the deformation at piezoresistive beams and the strain direction.  (b) Circuit diagram of the modified bridge network. The numbers denote the electric pads on a chip. Resistors of $470\ \mathrm{\Omega}$ were used as the external resistors. The output voltage was amplified by a low-noise differential preamplifier (SA-400F3, NF Electronic Instruments).}
\label{fig2}
\end{figure}

However, on-chip bridge circuits are ineffective for precise magnetic-torque measurement under strong magnetic fields because of the magnetoresistive effect. In contrast to the resistance change owing to strain in the membrane surface, the change that occurs because of magnetoresistance shows the same sign in all the resistors. Therefore, it gives a finite contribution to the output voltage unless all the resistors are perfectly matched.
To overcome this issue, we modified the on-chip aluminum interconnections as shown in fig.~\ref{fig1}(c). 
First, P$_2$ was disconnected from the bridge network by scraping the aluminum line. Then, Pads 2 and 3 were shorted. After this modification, the bias current flows only through P$_1$ and P$_3$; P$_2$ and P$_4$ thus become inactive. P$_1$ and P$_3$ are connected to the external resistors at room temperature to form a bridge network, as shown in fig.~\ref{fig2}(b).
The modified bridge circuit is actually equivalent to the one that was used for the measurement using a piezoresistive microcantilever~\cite{Ohmichi2002}, hence, it is possible to use the same experimental setup. The output voltage of the new configuration is given as follows:
\begin{equation}
\Delta V=\frac{R_{\mathrm{ext}}(\Delta R_3-\Delta R_1)}{2R_{\mathrm{ext}}+R_1 +R_3}I,
\end{equation}
where $R_{\mathrm{ext}}$ is the resistance of the external resistors. 
The different signs of the terms associated with P$_1$ and P$_3$ in the numerator show that the magnetoresistive effect is canceled out. However, this modification is very effective for torque measurement. 
Figure~\ref{fig2}(a) shows the relation between the magnetic torque and the strain exerted on P$_1$ and P$_3$.
In contrast to the isotropic deformation induced by gas adsorption, the strains exerted on P$_1$ and P$_3$ are in an opposite direction to each other. When P$_1$ is pulled upward and subjected to compressive strain, P$_3$ is pulled downward and subjected to tensile strain and vice versa. As a result, $R_1$ and $R_3$ show changes in the opposite direction and enhance the bridge output.

\begin{figure}[tb]
	\begin{center}
		\includegraphics[width=1\hsize]{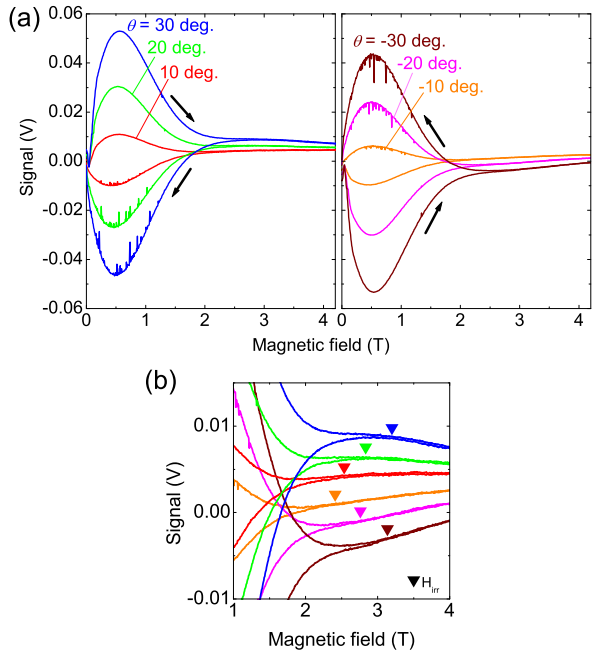}
	\end{center}
\caption{(a )Magnetic field dependence of the magnetic torque of $\kappa$-(BEDT-TTF)$_2$Cu(NCS)$_2$ sample. The left and right panel respectively show the data at positive and negative tilted angles. The black arrows show the sweep-up and sweep-down traces. (b) The same data around $H_{irr}$ indicated by the colored triangles.}
\label{fig3}
\end{figure}

To demonstrate the magnetic torque measurement using MSS, we chose $\kappa$-(BEDT-TTF)$_2$Cu(NCS)$_2$ as the first sample; this compound is a layerd organic superconductor whose transition temperature ($T_c$) of $10.4\ \mathrm{K}$~\cite{Ishiguro}.
The single-crystalline sample used in this study has a platelet shape with dimensions of $\sim0.3\times0.3\times0.13\ \mathrm{mm}^3$.
It was attached to the center of membrane using small amount of epoxy glue (Araldite$^{\textregistered}$ Rapid-AR-R30, Nichiban Co., Japan); the use of silicon grease in such cases is not recommended because its adhesive strength is not sufficient to hold the sample on the membrane at a low temperature. 
The sample was aligned such that the large flat surface was in contact with the membrane. In this case, the normal to the membrane corresponds to the stacking direction of layers. Then, the MSS chip was set on a vertically rotating stage for angle-dependent measurements. 
To maximize the strain, the rotation axis should be perpendicular to the direction connecting the active pair of piezoresistive beams. 
The bias current was set as $I=0.6\ \mathrm{mA}$.
The measurements were taken in $^4$He superfluid to suppress temperature increase that occurs owing to Joule heating of resistive paths.

Figure~\ref{fig3}(a) shows the magnetic field dependence of the magnetic torque at $T=1.5\ \mathrm{K}$.
$\theta$ is the angle between the magnetic field and the direction normal to the membrane. The hysteresis between sweep-up and sweep-down traces is caused by vortex pinning, which is the characteristic of the mixed state of type-I\hspace{-.1em}I superconductors.
As the magnetic torque $\bm{\tau} = \bm{M}\times \bm{H}$ ($\bm{M}$ is the sample magnetization, and $\bm{H}$ is the magnetic field), the amplitude of the hysteresis grows when $\theta$ increases. 
The sign of $\Delta V$ changes across $\theta=0\ \mathrm{deg.}$, which indicates the inversion of the stress direction.
The traces at positive and negative $\theta$ values are nearly symmetrical; however, there is a small difference in the hysteresis amplitudes owing to the misalignment of the rotating stage. A spike-like feature was observed only in sweep-down measurement. This is considered to be related to a flux jump~\cite{Mola2001, Sasaki1998}.

It should be mentioned that the upper critical field for the fields perpendicular and parallel to the superconducting layers are estimated to be $\mu_0 H_{c2}^{\perp}>5\ \mathrm{T}$ and $\mu_0 H_{c2}^{\parallel}>50\ \mathrm{T}$~\cite{Ishiguro, Lang1994}.
The magnetic fields at which the hysteresis disappear correspond to the irreversibility field $H_{irr}$. 
The magnetization process is not affected by vortex pinning above $H_{irr}$ because the thermally activated motion of the vortex lattice is dominant to the pinning potential.
Figure~\ref{fig3}(b) shows the expanded view of fig~\ref{fig3}(a) around $H_{irr}$.
The negligible drift shows the excellent long-time stability of the MSS. 
This enabled us to determine $H_{irr}$.
Note that $H_{irr}$ shifts to larger magnetic field as $\theta$ increases.
Such $\theta$ dependence is characteristic of two-dimensional superconductors.
All observed behaviors were consistent with the previous magnetic torque measurements using a microcantilever~\cite{Mola2001, Sasaki1998}.

\begin{figure}[tb]
	\begin{center}
		\includegraphics[width=0.9\hsize]{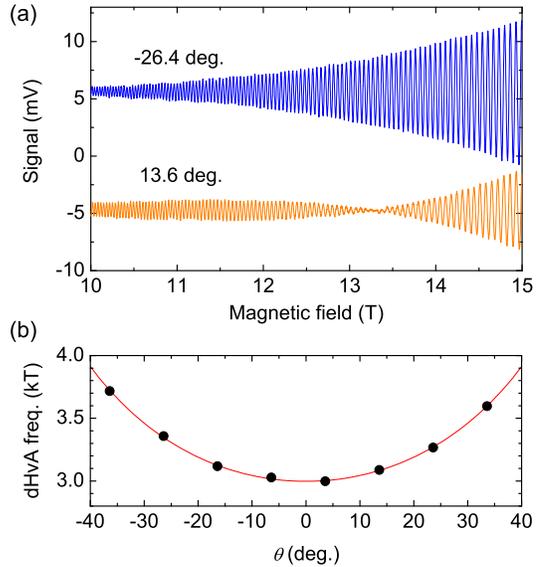}
	\end{center}
\caption{(a)dHvA oscillations of Sr$_2$RuO$_4$ at $\theta=13.6\ \mathrm{deg.}$ and $26.4\ \mathrm{deg.}$. The original misalignment $3.6\ \mathrm{deg.}$ was calibrated after the fitting in (b). (b) dHvA frequency as a function of $\theta$. The red solid line shows a fit to $F(0)/\cos(\theta)$ ($F(0)=3.00\ \mathrm{kT}$). }
\label{fig4}
\end{figure}

Next we performed dHvA oscillation measurement for the single crystal of the layerd oxide superconductor Sr$_2$RuO$_4$ ($T_c=1.5\ \mathrm{K}$) to check whether MSS can be used under higher magnetic field without a decreasing in its sensitivity.
Similar to the $\kappa$-(BEDT-TTF)$_2$Cu(NCS)$_2$ sample, this sample has a platelet shape ($\sim0.35\times0.35\times0.14\ \mathrm{mm}^3$) and the normal to its large surface is in the stacking direction of highly conductive layers.
For this experiment, we used the same MSS chip used in the first experiment. 
Araldite glue can be removed by soaking in acetone for a few tens of minutes; thus, the same chip can be used repeatedly. 
Moreover, no degrading effect was observed during this removal process.

Figure~\ref{fig4}(a) and \ref{fig4}(b) show the dHvA oscillation and oscillation frequencies of Sr$_2$RuO$_4$ at different $\theta$ values.
Clear oscillations were observed, and the difference in the shapes of the envelope originates from the warping of the Fermi surface.
The beating effect from multiple extremal orbits on a warped Fermi surface was observed at $\theta=13.6\ \mathrm{deg.}$.
In contrast, the amplitude monotonically increases at $\theta=-26.4\ \mathrm{deg.}$; at this angle, this material behaves as if it was an ideal two-dimensional system~\cite{Ohmichi2008}. 

As Sr$_2$RuO$_4$ has a two-dimentional cylindrical Fermi surface, the dHvA frequency follows the following equation: $F(\theta)=F(0)/\cos (\theta)$.
The fitting yielded $F(0)=3.00\ \mathrm{kT}$.
It has been reported that the Fermi surfaces are composed of three independent branches, namely $\alpha$, $\beta$, and $\gamma$, which correspond to dHvA frequencies of 3.05, 12.8, and 18.5 kT, respectively~\cite{Mackenzie1996, Ohmichi1999}.
Therefore, the observed oscillations are concluded to originate from the $\alpha$ branch.

The above experimental results ensure that MSS can be used as an alternative tool to microcantilevers.
We consider that a large membrane area is a distinct advantage of the MSS.
In the case of microcantilever, the mounted sample should be as small as $w^3$, where $w$ is the width of the cantilever beam (for PRC-400, $w=50\ \mathrm{\mu m}$) to avoid a mechanical damage.
In addition, it has been reported that a large sample size causes some issue, e.g., signal distortion~\cite{Ohmichi2002}.
On the other hand, since the MSS can hold a larger sample, the signal-to-noise ratio can be enhanced by increasing the sample volume. 
The samples used in the above experiments are indeed much larger than any samples that we have ever used in measurements using microcantilever. 
Nevertheless, the required sample size is still in the submillimeter scale, which is sufficient for practical use.

For existing users of PRC-series cantilever, Joule heating is likely a concern at much lower temperature than 1.5 K. The resistance values of piezoresistive paths of MSS is about ten times larger than those of PRC-series cantilever ($\sim 500\ \mathrm{\Omega}$). Therefore, the bias current should be $1/\sqrt{10}$ times as small as the value in the measurement using PRC cantilever. However, this does not mean the decrease of the signal-to-noise-ratio because the output signal is proportional to $\Delta R_i I$. For more quantitative comparison of the torque sensitivity of the two devices, we need the value of the torsional spring constant of MSS, which is unfortunately still unclear now. As for the point force at the center of the membrane, the spring constant is roughly estimated to be several to several tens of N/m~\cite{Akiyama}.

Finally, we mention other applications of MSS magnetometry.
One is the measurement under pulsed magnetic field~\cite{Ohmichi2002}.
The high eigenfrequency is required for the measurement that require fast response time.
The MSS has an eigenfrequency of $\sim 35\ \mathrm{kHz}$, which is almost the same as that of PRC-400, which has been used under pulsed magnetic fields.
The application of MSS to torque- and force-detected ESR spectroscopy is also promising~\cite{HT2015,OhmichiJIB,Ohmichi2016}.
In force-detected ESR, the on-chip bridge circuit, which was modified herein for torque magnetometry, would still be effective because the influence of the magnetoresistive effect can be eliminated by combining with the light modulation technique.

In conclusion, we developed a new method for torque magnetometry using a commercially available MSS chip. By modifying the on-chip aluminum interconnections, we achieved a stable operation and high sensitivity at a low temperature and under high magnetic fields. In addition, it is possible to use the same experimental setup that we used in the measurement using a piezoresistive microcantilever. Therefore, the MSS can be an alternative tool to self-sensitive microcantilevers.

The authors thank T. Akiyama (NANOSENSORS$^{\mathrm{TM}}$.) for having fruitful discussions and Y. Maeno and H. Anzai for supplying high-quality single crystals. This study was partly supported by a Grant-in-Aid for Young Scientists (B) (16K17749), Grant-in-Aid for Scientific Research (B) (No. 26287081) from the Japan Society for the Promotion of Science, and the Asahi Glass Foundation.

\end{document}